# Wobbling Ancient Binaries – Here Be Planets?

Jonathan Horner [1,2], Robert Wittenmyer [1,2], Tobias Hinse [3], Jonathan Marshall [4] and Alex Mustill [4]

[1] *School of Physics, University of New South Wales, Sydney, NSW 2052, Australia*
[2] *Australian Centre for Astrobiology, University of New South Wales, Sydney, NSW 2052, Australia*
[3] *Korea Astronomy and Space Science Institute, 776 Daedeokdae-ro Yuseong-gu 305-348 Daejeon, Korea*
[4] *Departamento de Física Teórica, Facultad de Ciencias, Universidad Autónoma de Madrid, Cantoblanco, 28049 Madrid, Spain*

**Summary:** In the last few years, a number of planets have been proposed to orbit several post main-sequence binary star systems on the basis of observed variations in the timing of eclipses between the binary components. A common feature of these planet candidates is that the best-fit orbits are often highly eccentric, such that the multiple planet systems proposed regularly feature mutually crossing orbits - a scenario that almost always leads to unstable planetary systems. In this work, we present the results of dynamical studies of all multiple-planet systems proposed to orbit these highly evolved binary stars, finding that most do not stand up to dynamical scrutiny. In one of the potentially stable cases (the NN Serpentis 2-planet system), we consider the evolution of the binary star system, and show that it is highly unlikely that planets could survive from the main sequence to obtain their current orbits - again casting doubt on the proposed planets. We conclude by considering alternative explanations for the observed variation in eclipse timings for these systems.

## Introduction

In the last two decades, a plethora of planets have been proposed orbiting a wide variety of stars. Depending on which of the two main online exoplanet databases you follow, the current tally of confirmed exoplanets stands at either 732 (http://exoplanets.org; [1]) or 992 (http://exoplanet.eu)[1]. Of those planets, only a handful (~1%) have been detected directly, through direct imaging (e.g. [2][3]) – the rest have been found through indirect means, such as the radial velocity (e.g. [4][5][6]) and transit (e.g. [7][8][9]) techniques. Although the bulk of the discoveries are made through these indirect means, it has still been possible to learn a great deal about the variety and nature of exoplanets – with recent work revealing the composition of exoplanetary atmospheres (e.g. [10][11]), the variety of exoplanet orbits (e.g. [12][13]) and the complexity of multiple planet systems (e.g. [14][15]).

The majority of those planets orbit single stars in the prime of their lives – main sequence stars (e.g. [16][17]). Fewer have been found around evolved stars – stars who have completed their main sequence life and swollen to become giants (e.g. [18][19]). However, in recent years, a number of planets have been proposed to orbit post-common envelope binary star systems[2] (hereafter PCEBs), on the basis of observed variations in the timing of eclipses between the two components of the binary system (as discussed in detail in e.g. [20]). Such planets, if they do indeed exist, would provide fascinating new insights into the planet formation process around binary stars, as well as the likelihood of planets surviving the post-main sequence evolution of their host stars. As such, it is clearly important to examine whether those planets are truly all that they seem to be. In most cases, PCEBs have been proposed to host two planets, often moving on surprisingly eccentric orbits. As a result, we have carried out a series of dynamical studies investigating the interaction of the proposed planets. Our work serves a dual purpose – first, it allows us to determine whether the candidate planets are dynamically feasible (that is, that their orbits are not catastrophically unstable). Second, our work can provide better constraints on the orbits of the planets in a given system (e.g. [15][21]), thereby allowing better characterisation of multiple planet systems.

---

[1] Exoplanet numbers taken from http://exoplanets.org and http://exoplanet.eu on 4[th] October 2013.

In this work, we summarise the results of our dynamical studies of all proposed multiple-planet systems orbiting PCEBs. In the next section, we briefly describe our methodology, before continuing to discuss the various systems studied. We follow that discussion with a brief summary of our subsequent investigations that go beyond pure dynamics, in order to better ascertain the true nature of the signals presented by the observed PCEBs, before drawing our conclusions in the final section.

## Dynamically Testing Exoplanet Systems

In order to test the dynamical stability of multiple planet exoplanetary systems, we follow a well-established routine (e.g. [15][22]). We use the Hybrid integrator within the n-body dynamics package MERCURY [23] to run a large suite of simulations following the evolution of the planets within the system for a period of 100 Myr. We hold the initial orbit of the best constrained planet fixed at the best-fit solution, and vary the initial orbit of the second candidate planet across solutions spanning the full ±3σ uncertainty range in orbital semi-major axis ($a$), eccentricity ($e$), argument of periastron ($\omega$) and mean anomaly ($M$). The planetary systems created in this way are then integrated forward in time for a period of 100 Myr, or until the planets therein collide with one another, are flung into the central star, or are ejected from the planetary system as a result of their mutual interactions. Should the system destabilise in this manner, the time at which this occurs is recorded.

---

[2] PCEBs are tightly bound evolved binary systems, which feature a white dwarf primary and a main sequence secondary star. These binaries were once more widely separated, but when the primary star moved off the main sequence, it expanded rapidly, to become a red giant star. At some point during its time on the Red Giant Branch or, slightly later, on the Asymptotic Giant Branch of the Hertzsprung-Russell diagram, the primary swelled to such an extent that it overflowed its Roche lobe, resulting in the two stars becoming surrounded by a common envelope of material. This began a period known as the "common envelope phase", during which the secondary star spiraled inwards until the loss of mass by the primary was completed, leaving it as a white dwarf or subdwarf B star. Where once the two stars orbited one another with a period of tens or hundreds of days, they now have a mutual orbital period of just a couple of hours. PCEBs frequently evolve to become "Cataclysmic Variable stars", with significant quantities of material being transferred from the secondary to the primary through an accretion disk, resulting in significant variability and even infrequent nova outbursts from the system. For more information, we direct the interested reader to [46], and references therein.

In this manner, we are able to create detailed maps of the dynamical stability of the planetary system in question, as a function of the initial semi-major axis and eccentricity of the orbit of the planet with the least tightly constrained orbital elements. We can then assess whether the planetary system is dynamically feasible and, if so, determine whether additional constraints can be placed on the orbits of the planets therein on the basis of the dynamics. For more details on our methodology, we direct the interested reader to [24], and references therein.

## Four Unstable Systems

Of the six PCEBs for which multiple companions have been proposed[3], four feature orbits for the candidate companions that cross one another. Mutually crossing orbits are typically dynamically unstable on short timescales, unless the objects moving on those orbits are protected from close encounters by the influence of mutual mean-motion resonance (as is the case for the Jovian and Neptunian Trojans (e.g. [26][27]) and the Plutinos [28] in our Solar system). The orbital solutions proposed for two of these systems (HW Virginis [29] and QS Virginis [30]) are presented in Figure 1 to illustrate the extreme architectures put forward in the discovery works.

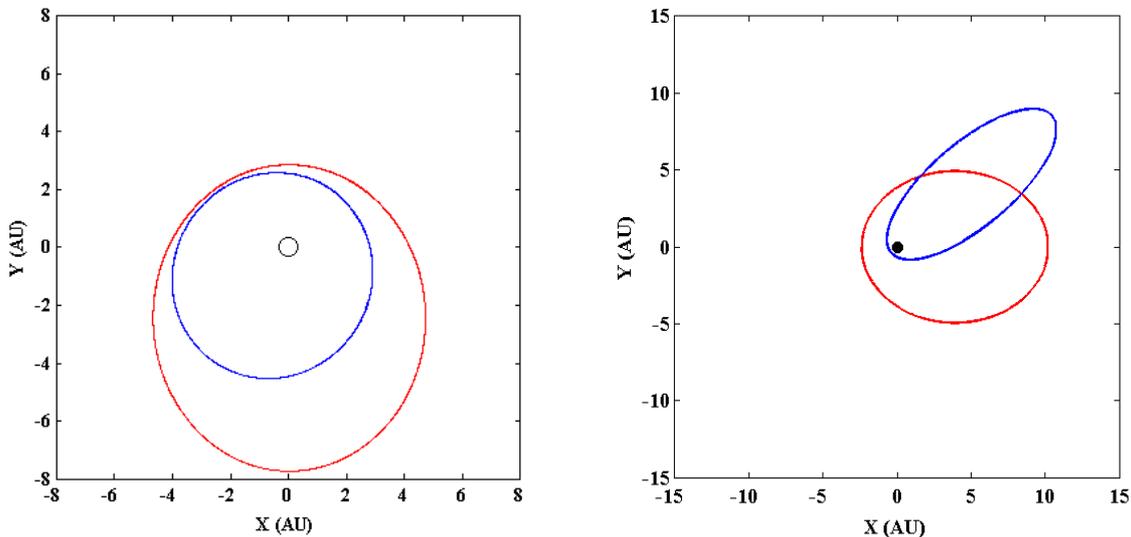

*Figure 1: The best-fit orbits proposed for the candidate planets around the PCEBs HW Virginis (left, [29]) and QS Virginis (right, [30]). Both scenarios feature orbits that approach one another closely. In the case of QS Virginis, the orbits have eccentricities reminiscent of the comets in our Solar system. For HW Virginis, although the best-fit orbits do not cross one another at the current epoch, the great majority of architectures permitted within the uncertainties on the orbital solution are mutually crossing.*

Given the extremely unusual architectures proposed for the candidate planets in these systems, it is clearly important to examine whether the systems could be dynamically feasible. We therefore studied each system in turn, to see whether any stood up to scrutiny. Table 1 presents the solutions proposed for the four systems in question in their discovery works, together with the uncertainties on the orbital parameters, and references to both the original work and our dynamical investigations. To illustrate the instability we observed for these systems, we present dynamical stability maps for HU Aquarii, HW Virginis, QS Virginis and V1828 Aquilae[4] (NSVS 14256825) in Figure 2.

---

[3] A seventh system, NY Vir, has one claimed companion, and the authors suggest there may be a second – but do not place any constraints on its orbital parameters [25]. As such, that system is not considered further in this work.

*Table 1: The best-fit orbital solutions for the four candidate PCEB companion systems that have proven to be dynamically unstable on astronomically short timescales.*

| System | Companion | Minimum Mass ($M_J$) | Semi-Major Axis (au) | Eccentricity | References |
|---|---|---|---|---|---|
| HU Aqr | b<br>c | 5.9±0.6<br>4.5±0.5 | 3.6±0.8<br>5.4±0.9 | 0.0<br>0.51±0.15 | [31][32][33] |
| HW Vir | b<br>c | 8.5±0.42<br>19.2±0.03 | 3.62±0.52<br>5.30±0.23 | 0.31±0.15<br>0.46±0.05 | [34][35] |
| V1828 Aquilae | b<br>c | 2.8±0.3<br>8.0±0.8 | 1.9±0.3<br>2.9±0.6 | 0.00±0.08<br>0.52±0.06 | [30][36] |
| QS Vir | b<br>c | 8<br>57 | 6.031±0.051<br>7.043±0.019 | 0.62±0.02<br>0.92±0.02 | [29][37] |

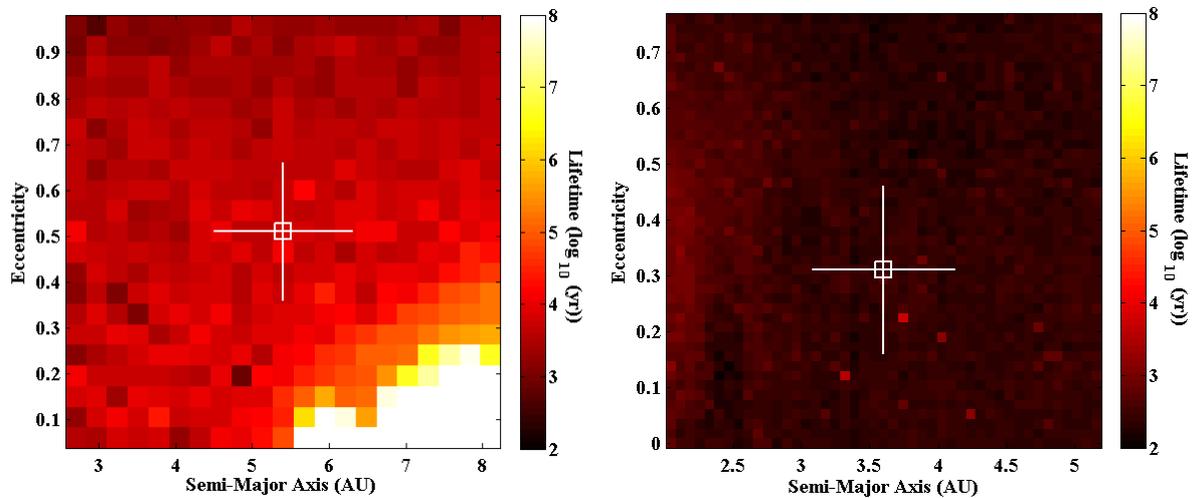

---

[4] It was brought to our attention during the review process for this paper that NSVS 14256825 was recently given the official variable star designation V1828 Aquilae, as part of the 80[th] name list of variable stars, details of which can be found at http://www.konkoly.hu/cgi-bin/IBVSpdf?6052 . As a result, we hereafter refer to it by its new designation.

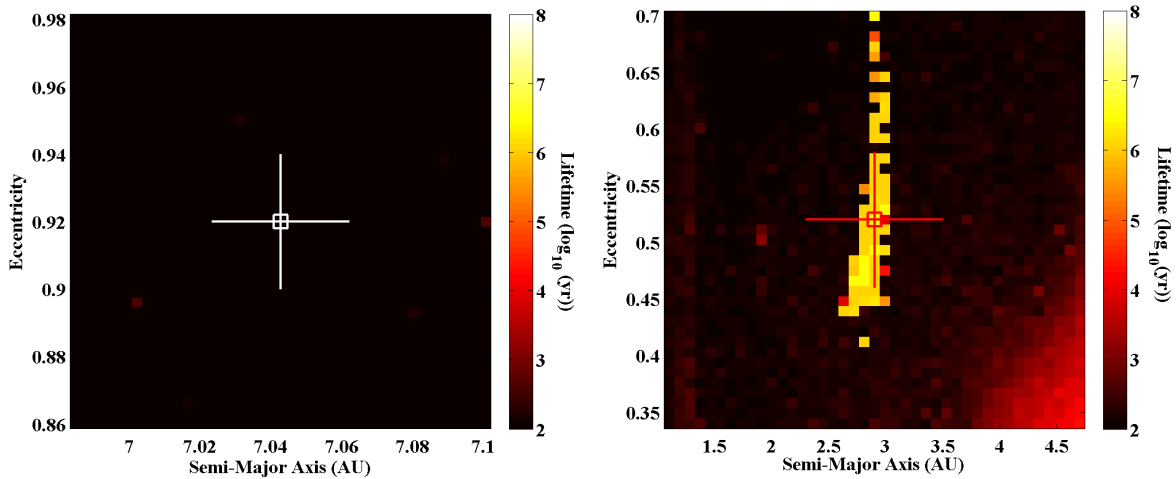

*Figure 2: Dynamical stability maps for HU Aquarii (top left), HW Virginis (top right), QS Virginis (lower left) and V1828 Aquilae (lower right). In each case, the stability is mapped as a function of the semi-major axis (a, in au) and eccentricity (e) of the companion with the most poorly constrained orbital elements. The colour at each (a-e) location shows the mean lifetime at that location, taken over all tested values of ω and M. The simulations for HU Aquarii were carried out at lower resolution than those for the other three systems, since at the time we had access to significantly less computational power. As a result, the lifetimes in each small square of that plot are the mean of 21 individual trials, whilst those for the other three are the mean of 75 unique trials.*

Each of the four systems detailed in Table 1 proves dynamically unstable on timescales of just a few hundred, or a few thousand years. Even in the case of V1828 Aquilae, the narrow region of enhanced stability features lifetimes of just ~ 1 Myr, far shorter than the estimated age and lifetime of the host stars. In each case, therefore, we conclude that the observed variations in the timing of eclipses between the two central stars are most likely not the result of unseen massive companions, although we acknowledge that the thin band of stability for V1828 Aquilae does suggest that, with future observations, it may be possible to find some dynamically plausible solutions for that system[5]. If such companions do exist, they must move on orbits drastically different to those proposed in the discovery works, and detailed in Table 1. It is interesting to note that three of the four systems detailed in Table 1 (HU Aqr, HW Vir and V1828 Aquilae) feature orbital semi-major axes in the ratio 1:1.5 (corresponding to orbital periods in a ratio ~1:1.84, i.e. nominal best-fit orbits just tighter than the mutual 1:2 mean-motion resonance between the proposed companions). The striking similarity between the orbital architectures proposed for these companions is interesting, and may perhaps suggest that the same physical process is responsible for the observed eclipse timing variations – if, indeed, those variations are real rather than an artefact of underestimated observational uncertainties[6].

---

[5] As further observations are obtained, the uncertainties on the fit to the orbit will decrease significantly, if the observed "wobble" is truly the result of massive companions. It is likely that such a decrease would be accompanied by some shift in the best fit values for the orbital rotation angles, which might we result in a "fine tuning" of the system towards a stable, resonant solution such as that hinted at by the narrow strip of stability shown in Fig. 2.

[6] We note that the problem of dynamically unfeasible companions claimed on the basis of eclipse timings between close binaries is not purely limited to PCEBs. The stars SZ Dra and RZ Dra, both of which are Algol type binaries, have also recently had massive companions proposed to orbit them on the basis of the timing of their mutual eclipses. As is the case for the four unstable systems described in this work, the candidate companions are proposed on

# The Stable Systems

The candidate planetary systems proposed around the PCEBs NN Serpentis ([42]) and UZ For ([43]) move on orbits that more closely resemble those of the planets in our own Solar system, and the bulk of known exoplanets. The orbits of the proposed companions have either low or moderate eccentricity and are relatively widely spaced, which in turn significantly reduces the likelihood of strong mutual perturbations on astronomically short timescales. The best-fit solutions proposed for these two candidate planetary systems are presented in Table 2.

In the case of NN Ser, the discovery work ([42]) proposed two potential architectures for the candidate planetary system, featuring companions trapped in mutual 2:1 and 5:2 mean-motion resonance. In our dynamical study of the system ([44]), we found that both these solutions would be dynamically stable for a wide range of plausible orbits for NN Ser (AB) d. However, we noted that the 5:2 resonant solution placed that planet on an orbit that was very close to a broad region of instability (see e.g. Fig. 4 in [44]). We therefore concluded that the 2:1 solution was the more likely of the two, when the stability of the orbits was considered in isolation. We present the best-fit orbits for that 2:1 solution, together with the dynamical stability map for the system, as the left-hand panels of Figure 3.

More recently, further observations of NN Ser have allowed the discoverers to significantly improve their orbital solution for the system ([45]), ruling out the 5:2 resonant solution. If the observed variations in the eclipses of NN Ser are the result of perturbations from massive companions, then it seems likely that those objects move on orbits that are trapped in mutual 2:1 mean-motion resonance, a conclusion consistent with our dynamical investigation.

The candidate planets orbiting UZ For move on orbits that are even more widely spaced than those in the NN Ser system, with orbital periods that place them just exterior to their mutual 3:1 mean-motion resonance. The closest analog to the two candidate planets in our own Solar system are the planets Saturn and Uranus, whose orbital periods differ by a factor of ~2.89 (compared to the ~3.05 difference between the periods of the UZ For companions). The orbital eccentricities of the candidate planets are almost identical to those of Saturn and Uranus – although their masses are significantly greater than the planets in our Solar system. Nevertheless, this comparison would instinctively suggest that the candidate planets should be dynamically stable on long timescales – an expectation that is borne out by the results of our simulations, as can be seen in the right hand panels of Figure 3. The great bulk of the tested solutions were dynamically stable, with the only broad region of instability lying beyond a semi-major axis of ~3.7 au for UZ For (AB) c. Orbits beyond this distance lie closer to that of UZ For (AB) b than the location of their mutual 2:1 mean-motion resonance, meaning that their instability is entirely to be expected (as was seen for the six systems discussed in the previous section).

*Table 2: The best-fit orbital solutions for the two candidate PCEB companion systems that have been shown to be dynamically feasible.*

| System | Companion | Minimum Mass ($M_J$) | Semi-Major Axis (au) | Eccentricity | References |
|---|---|---|---|---|---|

mutually crossing orbits, and prove dynamically unstable on astronomically short timescales. We direct the interested reader to [38] and [39] (SZ Dra) and [40] and [41] (RZ Dra) for more details on these systems.

| NN Ser | c | 6.91±0.54 | 5.38±0.20 | 0.0 | [42][44][45][46] |
|        | d | 2.28±0.38 | 3.39±0.10 | 0.20±0.02 |  |
| UZ For | b | 6.3 | 5.9±1.4 | 0.04±0.05 | [43] |
|        | c | 7.7 | 2.8±0.5 | 0.05±0.05 |  |

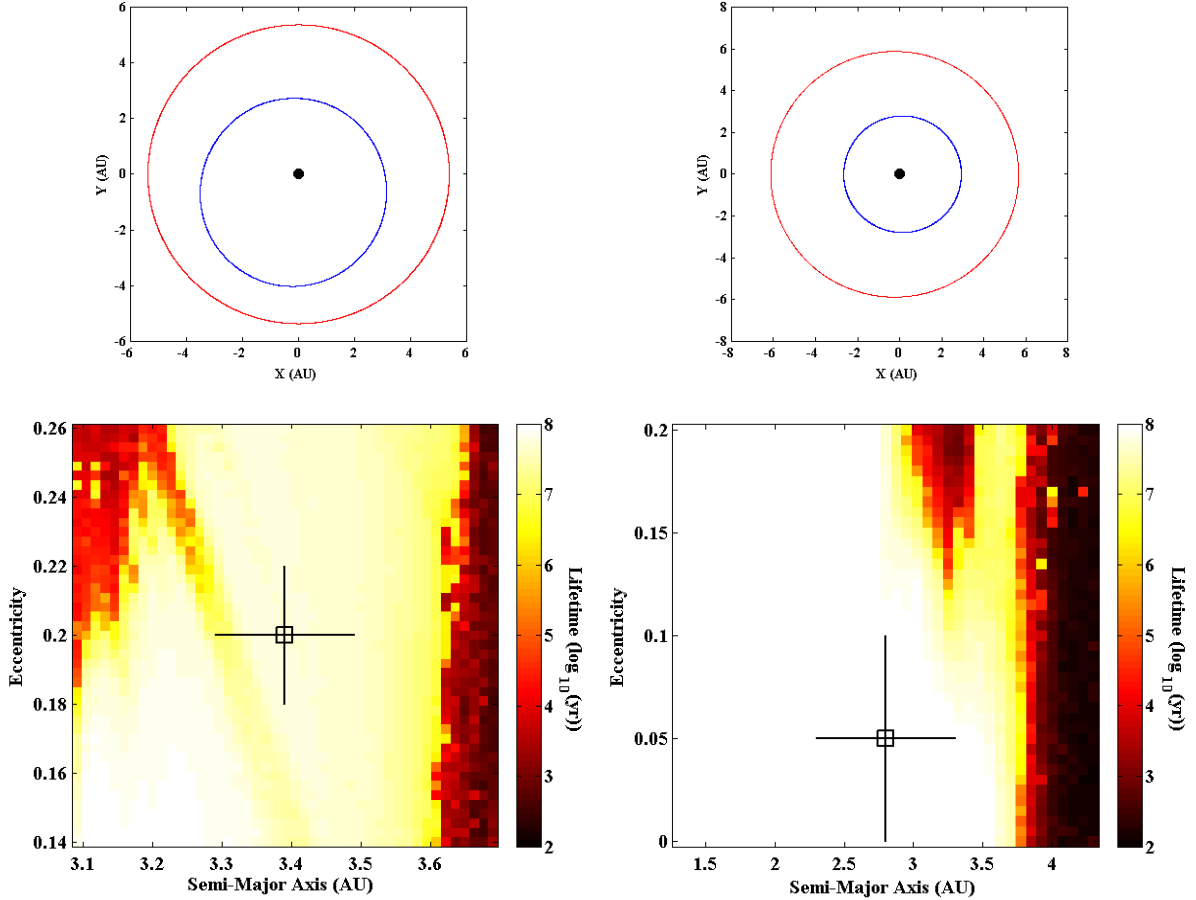

*Figure 3: The best-fit orbits (top) and dynamical stability maps (bottom) for the candidate planetary systems around NN Ser (left) and UZ For (right). For NN Ser, the stability map for the 2:1 resonant solution is shown. In both cases, the orbits are significantly more widely separated, and less eccentric, than those proposed for the other six candidate PCEB planetary systems. The best-fit solutions lie in broad regions of dynamical stability, in which the majority of solutions remained stable for the full duration of the 100 Myr simulations.*

## Circumbinary Planets and Stellar Evolution

Given that the planetary systems proposed orbiting NN Ser and UZ For stand up to dynamical scrutiny, it is interesting to consider whether the proposed planets could have formed at the same time as the host stars, before surviving the post-main sequence evolution of those two stars before acquiring their current orbits. We are therefore now in the process of studying the formation and evolution of these two candidate systems to see whether planets formed therein could survive the evolution of the system away from the main sequence to its current compact and dynamic state. Our study of the NN Ser system was recently completed ([46]), whilst that for UZ For has just begun.

In the case of the NN Ser system, we began with the assumption that the candidate planets formed with the binary star, and attempted to reconstruct a variety of initial binaries that could have evolved to give rise to the current configuration of NN Ser. Whilst the primary was still on the main sequence, the separation of the binary itself would have been much wider than that observed today, whilst the candidate planets would have moved on significantly smaller orbits. We used the binary star evolution code BSE ([47]) to follow the evolution of a large number of potential progenitor binaries through the primary's main sequence, red giant and asymptotic giant branch phases. Our modelling took account of the binary orbital changes due to their mutual tides and the common envelope phase, in which the two stars move within an envelope of material that consists primarily of the disrupted outer atmosphere of the giant primary. These effects combine, resulting in the orbits of the binary components decaying to the compact configuration observed today. In this way we identified the range of progenitor main sequence binary systems whose evolution renders them compatible with the current binary system. For more detail on our binary reconstruction procedure, we direct the interested reader to section 2 of [46].

We then investigated the stability on the main sequence of the potential progenitor systems identified through the stellar evolution modelling described above, when the binary's orbit was much wider and the planets' much tighter. The scale of the planetary orbits whilst on the main sequence was determined under the assumption that those orbits would have expanded adiabatically during the period of mass loss from the central binary as the primary evolved off the main sequence, as codified by Equation 1 in [46]. We followed the orbits of the bodies in the progenitor binaries with the MERCURY N-body code ([23]) over the primary's main sequence lifetime. Almost all the progenitor systems compatible with the evolution of the binary were dynamically unstable on very short timescales, typically of a few hundred or thousand years[7]. We therefore conclude that the candidate planets proposed to explain the variations in eclipse timing for NN Ser are not survivors from before the Common Envelope Phase. Either the objects formed from material that was not entirely ejected during the post-main sequence evolution of the system (which seems highly unlikely, see the discussion of [42], and references therein), or some other, non-planetary, explanation must be sought to explain the observed timing variations.

## Conclusion

Over the last few years, six multiple-companion systems have been proposed orbiting highly evolved binary star systems known as post-common envelope binaries (PCEBs). Those systems were all proposed based on periodic variations in the observed timing of eclipses between the two components of the evolved binary systems. We have shown that, of the six systems proposed, only two stand up to dynamical scrutiny – the rest are found to be catastrophically unstable on timescales of just a few hundred, or a few thousand years.

For the four dynamically unfeasible systems, it is clear that the proposed companions do not provide a reasonable explanation for the observed variations in the timing of the observed eclipses. Interestingly, a recent study of the circumbinary companions proposed to orbit the Algol-type binary RZ Dra has revealed that the uncertainties on the eclipse timings used in the discovery analysis may well have been significantly under-estimated ([41]). Indeed, in that work, we show that merely increasing the uncertainties by just ±5 seconds from their stated values is enough to make all evidence for the proposed companions disappear. Given that there are a number of factors that could deleteriously affect the precision with which the timing of mid-eclipse can be determined[8], it is not unreasonable to suggest that the uncertainties of these measurements may sometimes be underestimated. Indeed, we note that [31] adopt an uncertainty of just ±3.4 seconds for their eclipse timing observations, despite the fact that their observed eclipses display a large amount of variability and noise. Although we have not yet had chance to perform a similar analysis for the other unstable systems, it therefore seems reasonable to assume that the observations used in the discovery works may be victim to similar problems, and that therefore, once further observations of those systems are carried out, the evidence for unseen companions may disappear.

---

[7] The main sequence binaries corresponding to the current PCEB state of NN Ser featured binary semi-major axes between ~0.5 and 1.0 au, with the innermost planet residing at between ~1.04 and ~1.13 au. Such systems are inherently highly unstable, as a result of strong interactions between the planets and the secondary star. The observed instability is not unexpected – indeed, the instability can be predicted directly for all but the most compact main sequence binary configurations using the simple empirical expression given by equation 5 in [51].

In the case of the remaining two systems, NN Ser and UZ For, our dynamical simulations reveal that the candidate companions would move on orbits that are stable on long timescales. As a result, the existence of companions in those systems cannot be ruled out on the basis of their current dynamical evolution alone. As a result, we have begun a program of study investigating whether planets that formed with the host binary would be able to survive the binary's post-main sequence evolution before being emplaced on the orbits proposed around the PCEB. In the case of NN Ser, our study suggests that it is highly implausible that planets could survive the transition from the main sequence to the PCEB state for architectures that would result in the proposed planetary system ([46]). However, recent observations ([48]) have added further data to the archive for NN Ser, an analysis of which strengthens the conclusion that the observed periodic variation in eclipse timings for that system is a real effect – and that therefore the unseen companion hypothesis remains strongly supported by the observational data[9]. If the proposed planets do exist, then that system will prove a fascinating test-bed for models of exoplanet formation and evolution, and also for the fine details of post-main sequence evolution of close binary star systems.

## Acknowledgements

This research has made use of the Exoplanet Orbit Database and the Exoplanet Data Explorer at http://exoplanets.org. The work was supported by iVEC through the use of advanced computing resources located at the Murdoch University, in Western Australia. TCH gratefully acknowledges financial support from the Korea Research Council for Fundamental Science and Technology (KRCF) through the Young Research Scientist Fellowship Program, and also the support of the Korea Astronomy and Space Science Institute (KASI) grant 2013-9-400-00. JPM is supported by Spanish grant AYA 2011/26202. AJM is supported by Spanish grant

---

[8] Examples of such factors include the influence of star spots on the observed flux; [49], difficulties introduced by handling old data for which the time standard used has not been explicitly stated; [50], and the asymmetric and highly variable shape of eclipse light curves in PCEBs, as is well illustrated by Fig. 1 of [31].

[9] We note that, while this article was under review, [52] uploaded an article to the arXiv discussing a possible route by which the candidate planets orbiting NN Ser could have been formed from material ejected from the common envelope during the evolution of the system to become a PCEB – a model that will be tested by future observations of both NN Ser itself, and other PCEB systems.

AYA 2010/20630. The authors wish to thank the two anonymous referees of this work, whose comments led to changes that improved the flow and clarity of the final version.